\documentclass[a4paper]{article}

\usepackage{INTERSPEECH2022}
\usepackage{hyperref}
\usepackage{algorithm}
\usepackage{algorithmicx}
\usepackage{algpseudocode}
\usepackage{threeparttable}
\usepackage{makecell}
\usepackage[dvipsnames]{xcolor}
\usepackage{subfigure}
\usepackage{color} 
\usepackage{colortbl}

\title{CUSIDE: Chunking, Simulating Future Context and Decoding for Streaming ASR}
\name{Keyu An$^{1}$, Huahuan Zheng$^{1}$, Zhijian Ou$^{1}$$^{\dagger}$\thanks{$\dagger$ Corresponding author, also affiliated with Beijing National Research Center for Information Science and Technology, China. This work is supported by NSFC 61976122 and Tsinghua University - Meituan Joint Institute for Digital Life.}, Hongyu Xiang$^{2}$, Ke Ding$^{2}$, Guanglu Wan$^{2}$}
\address{
Speech Processing and Machine Intelligence (SPMI) Lab, Tsinghua University, China$^1$\\
Meituan, China$^2$}
\email{\{aky19,zhh20\}@mails.tsinghua.edu.cn, ozj@tsinghua.edu.cn, \\ \{xianghongyu, dingke02, wanguanglu\}@meituan.com}

\begin{document}
\ninept
\maketitle
\begin{abstract}
History and future contextual information are known to be important for accurate acoustic modeling. However, acquiring future context brings latency for streaming ASR. In this paper, we propose a new framework - \underline{{C}}h\underline{{u}}nking, \underline{{Si}}mulating Future Context and \underline{{De}}coding (CUSIDE) for streaming speech recognition. 
A new simulation module is introduced to recursively simulate the future contextual frames, without waiting for future context.
The simulation module is jointly trained with the ASR model using a self-supervised loss; the ASR model is optimized with the usual ASR loss, e.g., CTC-CRF as used in our experiments.
Experiments show that, compared to using real future frames as right context, using simulated future context can drastically reduce latency while maintaining recognition accuracy. With CUSIDE, we obtain new state-of-the-art streaming ASR results on the AISHELL-1 dataset.
\end{abstract}
\noindent\textbf{Index Terms}: Streaming ASR, Multi-task learning, Conformer.
\section{Introduction}
Recently, self-attention based neural networks such as Transformer~\cite{self-attention} and Conformer \cite{Conformer} have shown excellent performance in automatic speech recognition (ASR)~\cite{Conformer,Transformer-Transducer, Speech-Transformer}, particularly for being used as acoustic encoders to extract high-level representations from speech. Self-attention architectures capture temporal dependencies in a sequence by computing pairwise attention weights, and thus are superior in leveraging contextual information for acoustic encoders.
However, the full-sequence attention mechanism involves sequence-level fully-connected computation, and thus is unsuitable for streaming ASR, where each word must be recognized shortly after it was spoken. 
For streaming ASR, chunk-based self-attention networks, which use chunk-level input to control the latency, are adopted in many previous works~\cite{SCAMA,SAA,Online-CTC/Attention,CBP,CAT,Streaming-Transformer} and show better ASR performance when compared to other streaming schemes such as causal self-attention~\cite{MERL-DCN}.

In chunk-based latency controlled models, a certain number of left and right contextual frames are often spliced to each chunk, which is found to benefit the acoustic encoder and produce improved ASR performance~\cite{SAA,Online-CTC/Attention,CSC,CAT}. For example, Dong $et \ al.$~\cite{SAA} reported a 10\%$\sim$15\% relative character error rate reduction (CERR) on HKUST dataset when splicing left and right contextual frames to each chunk, and similar results are reported in other end-to-end models$~\cite{Online-CTC/Attention, CAT,Streaming-Transformer}$. However, the use of right contextual frames (i.e., look-ahead frames) needs to wait until the arrival of the right contextual frames, and thus brings additional latency.

\begin{figure}[!t]
	\centering
	\centerline{\includegraphics[width=7.5cm]{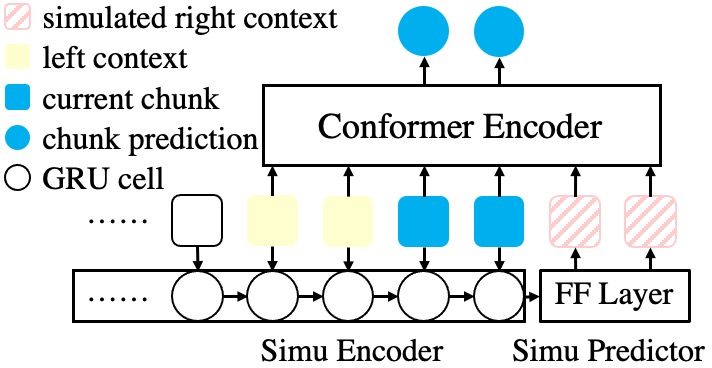}}
	\caption{The CUSIDE framework. 
	We split the utterance into non-overlapping chunks, and iteratively run simulation encoder and predictor to simulate a certain number of future frames. Current chunk is spliced with some history frames and the simulated future frames, and they are together fed to the acoustic encoder to produce the prediction for the current chunk.
	The simulation encoder, predictor, and acoustic encoder are implemented in our experiments as, but generally not limited to, GRU, feed-forward (FF) layer, and Conformer respectively.
	}
	\label{simu}
  	\vspace{-0.5cm}
\end{figure}

In this paper, we introduce a new framework - \underline{{C}}h\underline{{u}}nking, \underline{{Si}}mulating Future Context and \underline{{De}}coding (CUSIDE) for streaming ASR, which is illustrated in Figure \ref{simu}.
Inspired from predictive coding \cite{APC,CPC}, a new simulation module, consisting of lightweight simulation encoder and predictor, is introduced to recursively simulate the future contextual frames, based only on history frames. Thus, the ASR model does not need to wait for future context.
The CUSIDE framework can be easily applied to existing ASR models via multi-task training. 
The simulation module is jointly trained with the ASR model using a self-supervised loss; the ASR model is optimized with the usual ASR loss, e.g., CTC-CRF \cite{crf} as used in our experiments.
The CUSIDE framework can be further enhanced with several recently developed techniques for streaming ASR, such as weight sharing and joint training of a streaming model with a full-context model~\cite{dual-mode-asr}, chunk size jitter \cite{CAT,U2++} and stochastic future context \cite{MMTT} in training.
Experiments demonstrate that, compared to using real future frames as right context, using simulated future context can drastically reduce latency while maintaining recognition accuracy. With CUSIDE, we obtain new state-of-the-art streaming ASR results on the AISHELL-1 dataset \cite{aishell}.

The paper is organized as follows. 
Section 2 outlines related work. Section 3 details the components of CUSIDE. Experiments are shown in Section 4. Section 5 gives the conclusion. The code will be released upon the acceptance of the paper.


\section{Related work}
\subsection{Low latency ASR}
Building ASR models with low latency has been studied in conventional DNN-HMM hybrid systems~\cite{CSC}. Recently, three classes of end-to-end ASR models, namely CTC~\cite{ctc}, RNN-T~\cite{rnn-t} and attention-based encoder-decoder (AED) \cite{LAS}, have demonstrated excellent performance, which raises the demand to build streaming end-to-end ASR. 
\emph{For CTC and RNN-T}, both can use frame-synchronous decoding which involves no latency, thus the challenge for streaming ASR is to control the latency of the underlying acoustic encoder while maintaining recognition accuracy being close to a full-context model.
Streaming acoustic encoders can be built with uni-directional recurrent networks and causal self-attention networks, which do not use any future context, but with accuracy sacrifice.
Chunk-based acoustic encoders are attractive and adopted in many previous works, where bi-directional recurrent networks or fully-connected self-attention networks can be used within a chunk \cite{SAA,CAT,MERL-DCN}, realizing full-context utilization in a chunk.
Truncated self-attention and time restricted self-attention~\cite{Transformer-Transducer, resctricted-sa} are proposed to limit the contexts available for self-attention.
\emph{For AED}, the previously mentioned methods for low-latency acoustic encoders can be applied as well, but an additional challenge is that the decoding in AED is label-synchronous, which needs to attend to the entire sequence of representations extracted by the acoustic encoder.
To address this, many efforts have been made to convert full sequence soft attention into local attention~\cite{SCAMA, mocha,cif}. 

\subsection{Unifying streaming and non-streaming ASR} 
Recently, there has been a growing interest in unifying streaming and non-streaming ASR, with initial motivation to simplify the workflow of training and deploying streaming and full-context ASR.
Interestingly, dual-mode ASR~\cite{dual-mode-asr}, based on RNN-T, finds that weight sharing and joint training of a streaming model with a full-context model can benefit the low-latency and accuracy of streaming ASR.
U2~\cite{U2++}, based on hybrid CTC-AED, adopts dynamic chunk sizes in training, which are uniformly drawn from 1 to some maximum size. In this manner, different AED models, varying from causal attention to full attention, are jointly trained with shared weights.
Multi-mode Transformer Transducer~\cite{MMTT} proposes to consider stochastic future contexts during training, so that the trained model is capable to serve in various latency budget scenarios without significant accuracy deterioration.

Additionally, different two-pass decoding strategies have been used to unify streaming and non-streaming ASR, such as in Universal ASR~\cite{universal}, U2++ \cite{U2++} and RNN-T with Cascaded encoders ~\cite{cascaded}.
The main idea is that during inference, a streaming model is used first to generate intermediate results (features or hypotheses), which are then further processed by a non-streaming model. The encoders can be shared \cite{universal,U2++} or the decoders are shared \cite{cascaded}. 
In all the unified models mentioned above, the non-streaming model introduces extra parameters on top of the streaming model.

\subsection{Predictive coding}
The idea of predicting future frames has been explored in recent self-supervised speech representation learning such as autoregressive predictive coding (APC)~\cite{APC} and contrastive predictive coding (CPC)~\cite{CPC}. 
Beyond predicting the next few frames by exploiting the local smoothness of speech signal, it is shown in \cite{APC,CPC} that it is possible to learn more global structures in speech by neural networks so that the prediction can be as far as 200ms.
This success of predictive coding motivates the introduction of the simulation module in CUSIDE to simulate future context.
The difference is that these predictive coding methods adopt much larger encoders, which are trained on large amounts of unlabeled speech and used as speech representation learners for various downstream tasks such as speaker identification and speech translation. In CUSIDE, the purpose of predicting future is to provide context information for acoustic encoders for streaming ASR. 

\begin{figure}[t]
	\centering
	\centerline{\includegraphics[width=7.5cm]{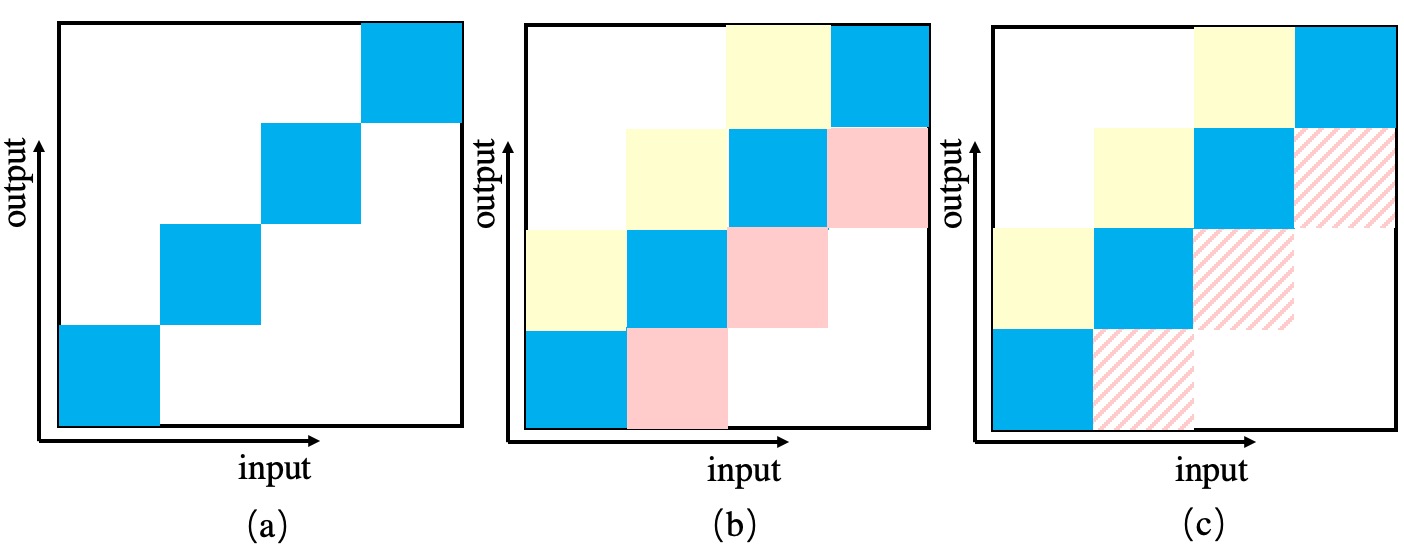}}
	\caption{The dependency map of different types of chunking used in inference. (a) Chunking w/o context. (b) Context sensitive chunking with left and right contexts, which are colored with yellow and red, respectively. (c) The proposed chunking with simulated future context. Colored [i, j] denotes whether output[j] depends on input[i]. White means there is no dependency. Dashed means that for output[j], input[i] is simulated.
	Here the sizes of left and right context are plotted to be equal to the sizes of the chunk, which is not necessary in practice.
	}
	\label{dep}
	\vspace{-0.6cm}
\end{figure}

\section{Method}
The CUSIDE framework consists of context sensitive chunking, simulating future context and decoding. 
It can be easily applied to existing ASR models via multi-task training as shown below.
Different neural network architectures (e.g., Bi-LSTM, Transformer and Conformer) can be used as acoustic encoders, and we can use different decoding methods, such as WFST based decoding for CTC or CTC-CRF~\cite{crf}, beam-search for RNN-T or AED~\cite{SCAMA,mocha}. To be specific, we adopt state-of-the-art Conformer encoder \cite{Conformer} for acoustic modeling and WFST based decoding for CTC-CRF in our experiments.

\subsection{Context sensitive chunking}
Full-sequence self-attention based encoders such as Transformer and Conformer require the entire speech utterance as input, thus unsuitable for streaming ASR.
To realize streaming ASR, we adopt context sensitive chunking (CSC) \cite{CSC,CAT} for the Conformer encoder, as illustrated in Figure \ref{dep}(b). Specifically, the entire utterance is firstly split into non-overlapping chunks. For each chunk, a certain number of frames to the left and right of the chunk are spliced as contextual frames. 
The spliced frames are collectively called a context sensitive chunk, which together is fed into the Conformer encoder. 
Note that the output from contextual frames is discarded at the output layer of the Conformer. Though with some extra computation, CSC is attractive and has been shown with promising results for streaming ASR \cite{CSC,CAT,SAA,Online-CTC/Attention}, due to its use of both past and future context. However, acquiring future context brings additional latency, as the model needs to wait until the right context frames come before it can produce current output during inference. To address this drawback, we propose CSC with simulated future context for low latency ASR.

\subsection{Simulating future context}
As illustrated in Figure \ref{simu}, in CUSIDE, the new arriving chunk is firstly encoded by a simulation encoder, which is a uni-directional GRU whose hidden state is initialized by the last hidden state from the previous chunk. The GRU hidden state $h \in \mathbb{R}^{\rm d_{GRU}} $ at the right boundary of the current chunk is then fed into a simulation predictor, which is a simple feed-forward (FF) layer, to simulate the right context frames $\widetilde{\bf X} = (\widetilde{\bf x}_1,\cdots,\widetilde{\bf x}_N)$, i.e., the log Mel-spectrograms:
\begin{equation} \label{simuNet}
\widetilde{\bf x}_i= h {\bf A}_i + b_i
\end{equation}
where $ \widetilde{\bf x}_i \in \mathbb{R}^{\rm d_{Mel}}, i=1,\cdots,N $, $N$ is the number the simulated right context frames, and $\rm d_{Mel}$ is the dimension of the log Mel-spectrograms. ${\bf A}_i$ is of shape ($\rm d_{GRU} $, $ \rm d_{Mel} $), and $b_i$ is of shape ($ \rm d_{Mel} $). 
In this paper, to reduce the overhead of parameters and computation due to the introduction of simulating future context, lightweight GRU and feed-forward layer are used as simulation encoder and predictor, respectively.

In both training and recognition, the simulated future context $\widetilde{\bf X}$ is spliced to the chunk with the left context, as shown in Figure \ref{dep}(c), which together are fed to the acoustic encoder as in usual CSC.
During training, the L1 loss between the simulated future frames $\widetilde{\bf X}$ and the real future frames ${\bf X} = ({\bf x}_1, \cdots, {\bf x}_N)$ is added to the ASR loss with a scaling factor $\alpha$. Thus, the simulation module is trained by the usual (supervised) ASR loss and the (self-supervised) simulation loss.


\subsection{Weight sharing and joint training with full-context model}
Inspired by dual-mode ASR~\cite{dual-mode-asr}, we train a single ASR model with shared weights for both streaming and non-streaming ASR. 
Notably, the forward calculation in the streaming mode of CUSIDE is actually in full-context within each context sensitive chunk, thus can be performed using exactly the same parameters and operations as in non-streaming ASR.
In contrast, the forward calculation in the streaming mode of dual-model ASR uses causal operators such as causal attention and convolution, which are different from non-causal operators used in the non-streaming mode. And the non-streaming model introduces extra parameters on top of the streaming model.
Moreover, we adopt chunk size jitter for different batches in training, which can improve the generalization capability of the streaming ASR model as shown in \cite{CAT, U2++}. Specifically, the chunk size for every batch is sampled from a uniform distribution $\mathcal{U}$(C-A, C+A), where C is the default chunk size and A is the dynamic range.

The whole training in CUSIDE as shown in Algorithm~\ref{algo} is in multi-task training, which involves three tasks -  simulation future context, streaming and non-streaming ASR.
We simulate future context for all chunks at the same time, and use the simulated frames and the ground-truth future frames (obtained by shifting the real frames backward) to calculate the simulation loss. 
For the streaming ASR loss, after chunk-wise forward calculation, we reshape the chunk-wise posterior predictions to construct the utterance-level posterior predictions, which are used to compute the streaming ASR loss. 
The non-streaming ASR loss is calculated with Conformer encoder as usual.

\begin{algorithm}
\footnotesize
\caption{Pseudocode of model training in CUSIDE.}
\label{algo}
\begin{algorithmic}
\State \textcolor{BlueGreen}{\# Requires: data$\_$loader, AcousticEncoder, SimuEncoder and SimuPredictor.}
    \For{X, y in data$\_$loader} \textcolor{BlueGreen} {\# Load the speech feature X and the label y.}
        \State non$\_$streaming$\_$pred $=$ AcousticEncoder(X) \textcolor{BlueGreen} {\# Perform full-context recognition.}
        \State non$\_$streaming$\_$loss $=$ asr$\_$loss(non$\_$streaming$\_$pred, y) \textcolor{BlueGreen} {\# Calculate utterance-level non-streaming loss.}
        \State
        \State simulated$\_$context = SimuPredictor(SimuEncoder(X)) \textcolor{BlueGreen}{\# Simulate the future context.} 
        \State simu$\_$loss = L1Loss(simulated$\_$context, Shift(X))
        \textcolor{BlueGreen}{\# Calculate simulation loss. Shift(X) outputs the ground-truth future context, by shifting X backward.}
        \State
        \State $\rm [chunk_k]_{1}^K$ = format$\_$chunk(X, simulated$\_$context) \textcolor{BlueGreen}{\# Construct K context sensitive chunks with simulated future context. K = $\rm \lceil len(X)/chunk\_size \rceil$. The chunk size can be randomly drawn. We use zero padding for the last chunk.}
        \State $\rm [chunk\_pred_k]_1^K$ = AcousticEncoder($\rm [chunk_k]_1^K$) \textcolor{BlueGreen}{\# Perform batch forward for all chunks and discard output from contextual frames.}
        \State streaming$\_$pred = format$\_$utt($\rm [chunk\_pred_k]_1^K$) \textcolor{BlueGreen}{\# Construct the utterance-level posterior prediction for the streaming model.}
        \State streaming$\_$loss = asr$\_$loss(streaming$\_$pred, y) \textcolor{BlueGreen}{\# Calculate utterance-level streaming loss.} 
        \State
        \State total$\_$loss = non$\_$streaming$\_$loss + streaming$\_$loss + $\alpha$*simu$\_$loss \textcolor{BlueGreen}{\# Compute total loss.}
        \State total$\_$loss.backward() \textcolor{BlueGreen}{\# Update parameters.}
    \EndFor
\end{algorithmic}
\end{algorithm}

\section{Experiments}
\subsection{Experiment settings}
We use the CTC-CRF based ASR Toolkit - CAT~\cite{CAT} to conduct the experiments. 
CTC-CRF \cite{crf} is a CRF with CTC topology, which eliminates the conditional independence assumption in CTC and performs significantly better than CTC.
Experiments are conducted on Aishell-1~\cite{aishell}, which consists of a total of 178 hours of labeled speech. The official lexicon ~\footnote{https://www.openslr.org/resources/33/resource\_aishell.tgz} is used.

All experiments use 80-dim filterbank features with 3-dim pitch features, which are extracted with a 25ms sliding window at a 10ms frame rate. We apply online mean and variance normalization to the input feature. 
3-fold speed perturbation and SpecAugment~\cite{specaug} are used for data augmentation. Unless otherwise specified, the acoustic model is a 12-layer Conformer, and the number of attention heads, attention dimension, and feed-forward dimension are 4, 256, and 2048 respectively. For simulating the future context, we use a 3-layer GRU with 256 hidden units, and 1 feed forward layer. The GRU and feed forward layer have about 5\% of the total model size ($\sim 37$M). Thus, the parameter overhead is negligible.

In training, we use the Adam optimizer and follow the Transformer learning rate scheduler~\cite{self-attention}. The learning rate will decay with a factor of 0.1 if the loss does not decrease on the validation set. We stop training when the learning rate is less than a given threshold. The final model averages the top 5 checkpoints with the best validation loss on the development set.  Gradient clipping is applied to stabilize the training. We further adopt a curriculum learning strategy in the first several training epochs, i.e., we sort the input utterances from short to long, as model training tends to be easier over short sequences. The scaling factor $\alpha$ for the simulation loss is set to 100. In decoding, a word-level 3-gram language model (LM) trained on the transcripts is used to build the decoding WFST.
\subsection{Experiment results}
The Character Error Rates (CERs) and average time costs per chunk when using different contextual information are shown in Table ~\ref{context}. From the table we observe that: 1) For the streaming ASR model, the lack of right context leads to a significant degradation in recognition accuracy (Exp 1 vs Exp 2, and Exp 6 vs Exp 7), and the accuracy degradation can not be compensated with more left context (Exp 1 vs Exp 7); 2) Using stochastic future context at training yields better performance at inference for the models without right context (Exp 4 and Exp 9) and with simulated right context (Exp 5 and Exp 10).  3) With stochastic future context, CUSIDE (Exp 5 and Exp 10) outperforms the model without right context (Exp 4 and Exp 9) consistently, with only a small overhead over parameter and time cost\footnote{2ms per chunk in our experiments when testing on a GeForce GTX 1080 GPU.},  especially when the CSC uses fewer left frames as history information. Notably, CUSIDE obtains equally good, or even better results compared with the model using real right context.
\begin{table}
\caption{CER and average time cost per chunk on AISHELL-1 test set. For the right context, $\rm [~]$ denotes that these frames are simulated. For all experiments, the chunk size is 400ms by default. ctx is shorthand for context.}
\vspace{-0.25cm}
\begin{threeparttable}
	\centering
	\scalebox{0.95}{
	\begin{tabular}{ccccc}
	\toprule
        \textbf{Exp} & \makecell[c]{\textbf{left ctx} \\ \textbf{(ms)}}   & \makecell[c]{\textbf{right ctx at} \\ \textbf{training (ms)}}  & \makecell[c]{\textbf{right ctx at} \\ \textbf{inference (ms)}}  & \textbf{CER}\\
        \midrule
        1 & 400 & 400 & 400 &  6.09\\
        2 & 400  & 0 & 0 & 7.27 \\
        3 & 400  & [400] & [400] & 7.28 \\
        4 & 400  & 400 or 0 & 0  & 7.15 \\
        5 & 400  & 400 or 0 or [400] & [400]  & \textbf{6.14} \\
        \midrule
        6 & 800  & 400 & 400 & 6.05  \\
        7 & 800  & 0 & 0  & 6.82 \\
        8 & 800  & [400] & [400]  & 6.73 \\
        9 & 800  & 400 or 0 & 0  & 5.94 \\
        10 & 800  & 400 or 0 or [400] & [400]  & \textbf{5.83} \\
		\bottomrule
	\end{tabular}}
\end{threeparttable}
\label{context}
\end{table}

The results of CUSIDE with chunk size jitter, weight sharing and joint training with a full-context model are shown in Table~\ref{ws_jt}. During training, the chunk size is randomly sampled from $\mathcal{U}$(200ms, 600ms) with C=400ms and A=200ms. The results demonstrate the effectiveness of these two training strategies. Notably, the performance of our streaming model is very close to the full-context model (5.47 vs 5.28). It is found that weight sharing and joint training also improve the non-streaming model (results in the parentheses). Presumably, this is because that the full-context model tends to overfit the training data, which could be mitigated by joint training with a chunk based model.
\begin{table}
\caption{Effect of chunk size jitter, weight sharing and joint training with a full-context model. 
The numbers in the parentheses are the full-context recognition results of the unified model.}
\vspace{-0.25cm}
\begin{threeparttable}
	\centering
	\scalebox{1}{
	\begin{tabular}{lc}
	\toprule
        \textbf{model configuration}  &  \textbf{CER} \\
        \midrule
        Exp 5 in Table~\ref{context} & 6.14  \\
    \quad + chunk size jitter in training & 5.96 \\
    \qquad + weight sharing \& joint training &  5.83 (4.98) \\
        \midrule
        Exp 10 in Table~\ref{context}  & 5.83 \\ 
    \quad + chunk size jitter in training  &  5.72 \\
    \qquad + weight sharing \& joint training & 5.47 (5.01) \\
    \midrule
    full-context model & 5.28 \\
		\bottomrule
	\end{tabular}}
\end{threeparttable}
\label{ws_jt}
\end{table}

We further compare the accuracy and latency of CUSIDE with other models from literature. Notably, some existing streaming models use two-pass decoding. For comparison, a 3-layer Transformer LM on the training transcripts in AISHELL-1 is used for rescoring. The results are shown in Table ~\ref{comparison}. CUSIDE obtains state-of-the-art results on AISHELL-1 with minimal overall latency.
\begin{table}
\caption{Comparison with other streaming models from literature on AISHELL-1 test set.  Following \cite{U2++}, the latency is defined as the chunk size plus the right context (if any). $\Delta$ is the additional latency introduced by rescoring the first-pass hypotheses, which is typically less than 100ms for a utterance.}
\vspace{-0.25cm}
\begin{threeparttable}
	\centering
	\scalebox{1}{
	\begin{tabular}{lcc}
	\toprule
    \textbf{model}  & \textbf{latency (ms)} &\textbf{CER} \\
    \midrule
    SCAMA~\cite{SCAMA} 
    & 600 & 7.39 \\
    MMA  narrow chunk ~\cite{MMA}
    & 960 &  7.5 \\
    MMA  wide chunk ~\cite{MMA}
    & 1920 &  6.6 \\
    HS-DACS Transformer~\cite{HS-DACS}
    & 1280 &  6.8 \\
    U2++~\cite{U2++}
    & 640 & 5.81 \\
    U2++ w/ rescoring~\cite{U2++}
    & 640 + $\Delta$ & 5.05 \\
    WNARS w/ rescoring~\cite{WNARS}
    & 640 + $\Delta$ & 5.22 \\
    \midrule
    CUSIDE & 400 + 2 & 5.47 \\
    \quad + NNLM rescoring & 400 + 2 + $\Delta$ & 4.79 \\
	\bottomrule 
	\end{tabular}}
\end{threeparttable}
\label{comparison}
\end{table}

\begin{figure}[!h]
  \centering
  \includegraphics[width=0.49\linewidth]{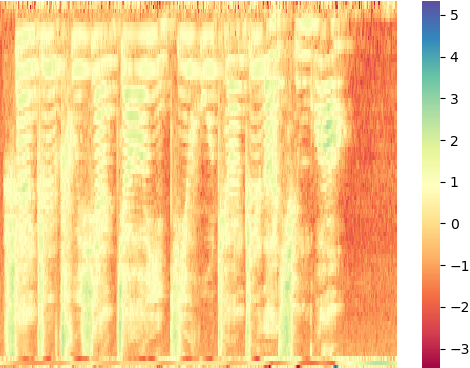}
  \includegraphics[width=0.49\linewidth]{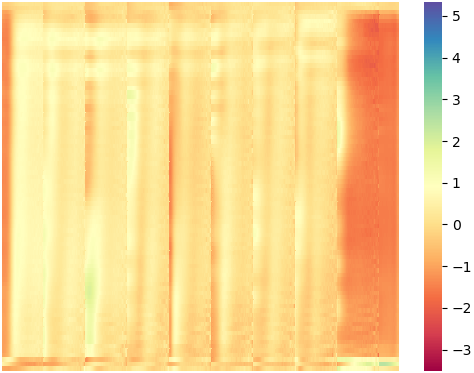}
  \caption{The real spectrogram (left, with mean and variance normalized) and the simulated spectrogram (right) for the utterance (BAC009S0764W0121) in the AISHELL-1 test set.}
  \label{fig:simu_vis}
\end{figure}

\subsection{Discussion}
To observe the quality of the simulated frames, we visualize the simulated log Mel-spectrogram in Fig~\ref{fig:simu_vis}. It can be seen that the simulated result roughly follows the spectrogram patterns, e.g., the different magnitudes of the speech and non-speech segments, but may not be able to capture finer details, e.g., the formant peaks in middle and high frequencies. As we only use limited data to train the simulation module, further improving the simulation module such as self-supervised pretraining on larger datasets would be an interesting extension of our work.

\section{Conclusions}
In this paper, we introduce the CUSIDE framework, which uses simulated future  context to improve the recognition accuracy of chunk-based streaming ASR.  With CUSIDE, we achieve new state-of-the-art streaming ASR results on AISHELL-1 in terms of accuracy and latency. While we use Conformer based CTC-CRF as our ASR model, our approach can be applied to other end-to-end ASR models such as RNN-T and attention based encoder-decoder, which will be interesting future work.

\bibliographystyle{IEEEtran}
\bibliography{mybib}

\end{document}